\documentclass[doc]{article}
\usepackage[utf8]{inputenc}
\usepackage{amsmath, amssymb}
\usepackage[a4paper, margin=1in]{geometry}
\usepackage{authblk}
\usepackage{graphicx}
\graphicspath{ {./images/} }
\usepackage{caption}
\usepackage{subcaption}
\usepackage[table,xcdraw]{xcolor}
\usepackage[parfill]{parskip} 
\usepackage{apacite}
\usepackage{float}
\DeclareMathOperator*{\argmax}{argmax}

\title{A Bayesian Dynamical System Model \\ of Joint Action and Interpersonal Coordination}
\author[1,*]{Andrew Jun Lee}
\author[2]{Grace Qiyuan Miao}
\author[2]{Rick Dale}
\author[3]{Alexia Galati}
\author[1,4]{Hongjing Lu}
\affil[1]{Department of Psychology, University of California, Los Angeles}
\affil[2]{Department of Communication, University of California, Los Angeles}
\affil[3]{Department of Psychological Science, University of North Carolina at Charlotte}
\affil[4]{Department of Statistics, University of California, Los Angeles}
\affil[*]{Corresponding Author: andrewlee0@ucla.edu}

\begin{document}
\date{}
\maketitle

\begin{abstract}
\noindent
Successful teamwork depends on interpersonal dynamics, the ways in which individuals coordinate, influence, and adapt to one another over time. Existing measures of interpersonal dynamics, such as CRQA, correlation, Granger causality, and transfer entropy, typically capture only a single dimension: either the synchrony/coordination or the direction of influence between individuals. What is missing is a psychologically meaningful representation that unifies these dimensions and varies systematically with behavior. We propose the \textit{context matrix} as one such representation. The context matrix, modeled within a linear dynamical system, has psychologically interpretable entries specifying how much each individual’s current behavior is attributable to their own versus every other group member’s past behaviors. Critically, these entries can be distilled into summary features that represent synchrony and directional influence. Evidence for the context matrix as psychologically meaningful is provided in two steps. First, we develop a sequential Bayesian model that infers context matrices from timeseries data and show that it accurately recovers them in noisy simulations. Second, applying the model to human eyetracking data, we demonstrate that summary features of the inferred context matrices capture expected task-based differences in interpersonal dynamics (or lack thereof), predict task accuracy in psychologically reasonable ways, and show some correspondence with existing measures (CRQA and Granger causality). We conclude by situating the context matrix within a broader agenda for modeling interpersonal dynamics in joint action.
\end{abstract}

\section{Introduction}
Humans regularly coordinate in teams to achieve what individuals cannot accomplish alone. Intuition tells us that teams succeed by pooling multiple perspectives, a view captured by the adage ``two heads are better than one," but multiplicity of minds is no guarantee of success (e.g., \citeNP{bahrami2010optimally}). Team performance depends not only on who is present but also their interpersonal dynamics: the ways individuals coordinate, influence, and adapt to one another \cite{mcgrath1984groups, arrow2000small}. History offers costly reminders of teamwork gone awry. The failed U.S. Bay of Pigs invasion of 1961 was spurred by political leaders who prioritized cohesion over critique. This imbalance fostered social pressure, poor communication, and the illusion of consensus---a suboptimal dynamic known as ``groupthink" \cite{janis1972victims}. 

While interpersonal dynamics are largely shaped by fixed characteristics like personality, they are also sensitive to the demands of the task at hand \shortcite{coco2018performance, galatietal}. For instance, individuals who typically engage as equals may shift to a leader-follower relationship when one partner possesses greater task expertise than the other \shortcite{goldstone2024emergence}. In a model-car construction task, \shortciteA{wallot2016beyond} found that behavioral synchrony (temporal correlation) increased when teams followed a shared building strategy, and greater synchrony predicted subjective satisfaction and objective performance. Similarly, dyads sharing visual gaze completed search tasks faster than those using only voice or no communication at all \shortcite{brennan2008coordinating}. In a simulated military task, teams told to reflect on and revise their strategy outperformed those that did not \shortcite{gurtner2007getting}.

\begin{figure}
\centering
\includegraphics[width=0.85\textwidth]{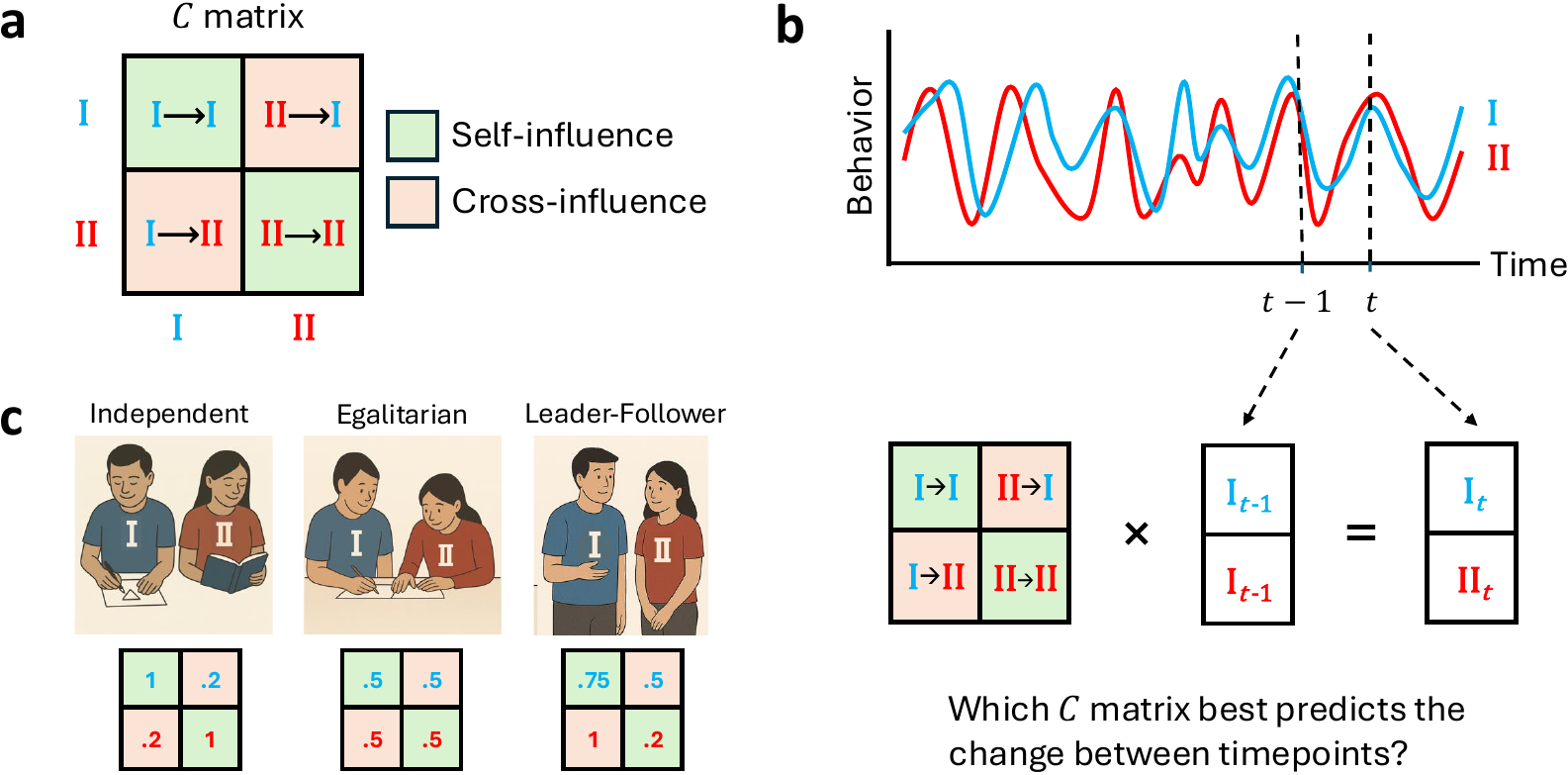}
\caption{Understanding the context matrix. (a) The context matrix is a quantitative description of a team's interpersonal dynamic. It consists of two core dimensions: self-influence (diagonal elements) and cross-influence (off-diagonal elements), where each element is a proportion. (b) The context matrix models behavioral change over time. The goal is to determine the context matrix that best predicts the change between timepoints. An individual's behavior at a given timepoint is predicted as the weighted sum of both members' behaviors from the previous timepoint, with weights given by the individual's corresponding row in the context matrix. (c) Many types of interpersonal dynamics can be expressed by the context matrix.}
\label{fig:intro_figureA}
\end{figure}

However, despite decades of research, the concept of interpersonal dynamics remains poorly formalized. One widely studied aspect of interpersonal dynamics is temporal synchrony, the degree to which individuals' behaviors align over time. Metrics such as cross-recurrence quantification analysis (CRQA) and its multidimensional variant (MdRQA) assess synchrony as a complex correlation between behavioral timeseries \shortcite{wallot2018analyzing, wallot2016multidimensional, konvalinka2011synchronized}. Spectral analyses, such as Fourier analysis, reflect synchrony by converting timeseries into frequencies, in which similarities of spectral peaks indicate shared behavioral rhythm and differences indicate complementary roles \shortcite{nalepka2019human}. Another aspect of interpersonal dynamics is the direction of influence (e.g., who is responding to whom). This can be measured as transfer entropy \shortcite{takamizawa2019transfer} or Granger causality \shortcite{chang2017body}.

Is there a metric that captures multiple facets of interpersonal dynamics in one interpretable framework? In this paper, we contend that the ``context matrix," introduced by our group \shortcite{miao2023mis}, is one such candidate. The context matrix is a compact representation of how individuals in an arbitrarily-sized group influence themselves and each other over time. Every possible pairwise influence is encoded: each entry specifies either a self-influence term (how much of one’s current behavior is explained by one’s own past behavior) or cross-influence term (how much is explained by a partner’s past behavior). For two people, this yields a $2 \times 2$ matrix where a row represents an individual and the two columns contain the self- and cross-influence terms. We model behavior as a linear dynamical system in which an individual's behavior at timepoint $t$ is a weighted sum of all individuals’ past behaviors at $t-1$, weighted according to a context matrix at $t$ (Figure~\ref{fig:intro_figureA}b). Linear dynamics is not the only possible dynamics, but provide a simple way to interpret each matrix entry as a causal weight linking past to present. For example, the matrix $[1, .2; 0, 1]$ means the first individual's behavior is 100\% of their previous behavior and 20\% of their partner's, while their partner's behavior is just 100\% of their own previous behavior. 

Despite its simplicity, the context matrix expresses a wide variety of interpersonal dynamics by simultaneously encoding synchrony and directional influence (Figure~\ref{fig:intro_figureA}c). Each self- and cross-influence term can be independently tuned, allowing for diverse combinations of self- and cross-influence, which produce varying patterns of synchrony and directional influence \shortcite{miao2023mis}. For example, high self-influence and low cross-influence yields independent behaviors (i.e., minimal synchrony), whereas low self-influence and high cross-influence produces synchronous behaviors. Furthermore, asymmetries in cross-influence capture leader–follower dynamics, where an individual mirrors their partner who remains unaffected by the individual. Small changes in these terms yield subtle variations: a leader with strong self-influence but little responsiveness appears unyielding, whereas one heavily influenced by a follower will lead in circles. 

But is the context matrix a psychologically accurate representation of interpersonal dynamics? In prior work, we generated timeseries data from a predefined context matrix according to the linear dynamics above \shortcite{miao2023mis}. These simulations assume that interpersonal dynamics follow a single context matrix that does not change over time. This stationary assumption may not hold in real team interactions, where interpersonal dynamics frequently shift and adapt as a task progresses. To establish the psychological validity of the context matrix and, more broadly, the utility of tasks as a core determinant of interpersonal dynamics, we provide evidence to satisfy two criteria: (1) it should be possible to \textit{accurately} infer the context matrices from behavioral timeseries, and (2) the inferred matrices should be \textit{useful} explanatory constructs of psychological phenomena in that, for example, they vary with task-based or individual differences.

\subsection{Current Study}
To satisfy both criteria, we need a model to infer context matrices from behavioral timeseries. We propose a sequential Bayesian model that estimates context matrices most likely to predict changes in behavior across adjacent timepoints. To do so, the model computes a \textit{posterior }distribution of possible context matrices, assigns each matrix a probability, and identifies the most probable candidate. A matrix's posterior probability is based on its likelihood and prior probabilities. The likelihood is computed from the error between the behavior it predicts (based on the previous time step) and the observed behavior. The prior is a ``smoothed" posterior distribution from the previous timepoint, called a temporal smoothness prior. The temporal smoothness prior assumes that interpersonal dynamics are unlikely to change dramatically over short periods \cite{kitagawa1996smoothness, gersch1991smoothness}. As the model proceeds through the timeseries and accumulates evidence, the posterior gradually favors context matrices that consistently predict the observed data.

While alternative inferential methods exist, we choose Bayesian inference for two reasons. First, Bayesian inference is a successful model of much of cognition, including concept learning, analogical reasoning, logical inference, physical reasoning, causal inference, belief updating, memory errors, visual and auditory perception, language learning, and social coordination (\textit{inter alia}: \shortciteNP{anderson1992explorations, chater2006probabilistic, frank2011computational, griffiths2010probabilistic, hawkins2023partners, lu2008bayesian, lu2016bayesian, tenenbaum2011grow, yuille2006vision}). We believe its broad applicability stems from two strengths: the capacity to (1) represent an indefinitely large space of candidate hypotheses (e.g., possible concepts, mappings, causes), including the correct one, and (2) encode inductive biases in the prior that guide inference toward that hypothesis. Our model captures these strengths by representing a wide distribution of context matrices, which become increasingly informative priors as time unfolds.

A second reason for embracing a Bayesian approach is its potential for novel theoretical integration. Cognitive science has long drawn a sharp distinction between rational and Bayesian frameworks on the one hand and dynamic and connectionist ones on the other, each grounded in different assumptions about representation, computation, and the nature of cognition \cite{dale2008possibility, jones2011bayesian, michaels2008direct, mcclelland2010letting}. This is a surprising tension considering that both frameworks embrace the fundamental importance of statistics in the environment and the probabilistic nature of cognitive agents that learn and adapt in that environment. Some have recently argued that integrating these broad traditions offers opportunities for a broader understanding of cognitive systems and the tasks they solve \shortcite{dale2022dissipation} and to have seemingly competing frameworks participate together as distinctive explanatory schemes that can unveil new aspects of an underlying cognitive phenomenon \shortcite{horst2022pluralism}. Our Bayesian model makes progress toward this goal by modeling a linear dynamical system (e.g., the context matrix as the state-transition matrix).

We begin by validating the model on synthetic data, showing that it successfully recovers known context matrices from simulated timeseries, even under high levels of noise (Criterion 1). We then apply the model to data from human tasks \shortcite{galatietal}, demonstrating that the inferred context matrices vary systematically across tasks (Criterion 2). We conclude that the context matrix is a compact and powerful approximation of interpersonal dynamics in real-world interaction.

\section{Modeling Framework}
\subsection{Sequential Bayesian Inference}
Consider two individuals, $\text{I}$ and $\text{II}$, performing a joint task. Both individuals' behaviors at time $t$ are represented as a matrix of behaviors with an arbitrary number of behavioral channels \textit{H}, such as speech or eye-gaze,  $$\textbf{b}_t =
\begin{bmatrix}
b_{t, \, \text{gaze}}^{\text{I}} & b_{t, \, \text{speech}}^{\text{I}} & \dots & b_{t, H}^{\text{I}} \\
b_{t, \, \text{gaze}}^{\text{II}} & b_{t, \, \text{speech}}^{\text{II}} & \dots & b_{t, H}^{\text{II}}
\end{bmatrix}\in \mathbb{R}^{2 \times H}.$$ The behaviors for a channel $h$ form a column vector of this matrix, $$
\textbf{b}_{t,h} = \begin{bmatrix}
b_{t,h}^{\text{I}} \\
b_{t,h}^{\text{II}} \\
\end{bmatrix} \in \mathbb{R}^2
$$ and the entire timeseries of behaviors is the three-tensor, $$\textbf{B} = \textbf{b}_{1:T} = [\textbf{b}_1,\dots,\textbf{b}_T] \in \mathbb{R}^{2 \times H \times T}.$$ Our goal is to infer a timeseries of context matrices $$C^* = [C^*_2, \dots, C^*_{t-1}, C^*_t, C^*_{t+1}, \dots]$$ in which each $C^*_t$ best explains the change in behaviors over adjacent timepoints $\textbf{b}_{t-1}\text{ and } \textbf{b}_t$, and starts with $C_2^*$ when two behaviors are observed. Formally, the context matrix $C^*_t$ that best explains the change $\textbf{b}_{t-1} \rightarrow \textbf{b}_t$ is the most probable in a distribution of possible matrices for time $t$ given the adjacent behaviors: $$C^*_t = \argmax_{C_t}\big[P(C_t|\textbf{b}_t, \textbf{b}_{t-1})\big].$$

We estimate the distribution $P(C_t|\textbf{b}_t, \textbf{b}_{t-1})$ as the posterior in sequential Bayesian inference, a framework that has successfully modeled causal learning from sequential data \cite{lu2016bayesian, abbott2011exploring}. 

In its standard form, sequential Bayesian inference conditions the posterior distribution of a hidden state (here, the context matrix), on the full history of behaviors: $$P(C_t|\textbf{b}_{1:t})$$ where $\textbf{b}_{1:t} = [\textbf{b}_1,\dots,\textbf{b}_t]$. Bayes' theorem states the posterior is proportional to the product of the likelihood and prior distributions: $$\underbrace{P(C_t \mid \textbf{b}_{1:t})}_{\text{Posterior}} \;\propto\;
\underbrace{P(\textbf{b}_t \mid C_t, \textbf{b}_{1:t-1})}_{\text{Likelihood}}
\;\cdot\;
\underbrace{P(C_t|\textbf{b}_{1:t-1})}_{\text{Prior}}.$$ Here, the likelihood is the probability of obtaining current behavior $\textbf{b}_t$ given some $C_t$ and the previous history $\textbf{b}_{1:t-1}$, and the prior is the probability of $C_t$ before observing the current behavior $\textbf{b}_t$. 

For tractability, modelers typically assume a first-order Markov property, which states that the posterior of a hidden state depends only on the immediately preceding behavior. We also assume this property, making the two forms equivalent, $$P(C_t|\textbf{b}_{1:t}) = P(C_t|\textbf{b}_t,\textbf{b}_{t-1})$$and allowing us to rewrite standard sequential Bayes as follows: $$P(C_t \mid \textbf{b}_{t}, \textbf{b}_{t-1}) \;\propto\; P(\textbf{b}_t \mid C_t, \textbf{b}_{t-1}) \;\cdot\; P(C_t|\textbf{b}_{t-1}).$$ The likelihood is now the probability of $\textbf{b}_t$ given some $C_t$ applied to just $\textbf{b}_{t-1}$ and the prior is conditional on only the previous timepoint $\textbf{b}_{t-1}$.

We define likelihood as the multivariate normal density function with input $\textbf{b}_t$ centered at predicted behavior $\hat{\textbf{b}}_t$, the behavior \textit{predicted} by $C_t$ from $\textbf{b}_{t-1}$: $$P(\textbf{b}_t|C_t,\textbf{b}_{t-1}) = \mathcal{N}\!\big(\textbf{b}_t \;\big|\; \hat{\textbf{b}}_t, \, \Sigma_\textbf{B} \big)$$ where $\Sigma_\textbf{B}$ is the matrix $[\text{var}(\textbf{B}^\text{I}) \,\, 0; \,\, 0 \,\, \text{var}(\textbf{B}^{\text{II}})]$ with the diagonal set to the variance of each individual's behavioral measures across entire timeseries, $\textbf{B}^{\text{I}}$ and $\textbf{B}^{\text{II}}$. Using the multivariate normal distribution as likelihood captures the intuition that predictions from a model should be close to the observed behaviors. We use the multivariate normal distribution for simplicity, but alternative distributions with non-normal characteristics could also be possible in some applications. 

When $\textbf{b}_t$ is a matrix with multiple behavioral channels (e.g., speech, eye-gaze), a likelihood is computed for each channel by setting the mean of the multivariate normal density function to the matrix column corresponding to the channel $\textbf{b}_{t, h}$. The likelihood for $C_t$ is then given by multiplying these separate likelihoods, $$P(\textbf{b}_t|C_t,\textbf{b}_{t-1}) = \prod_h^H P(\textbf{b}_{t,h}|C_t,\textbf{b}_{t-1, h}).$$

The predicted behavior $\hat{\textbf{b}}_t$ derived from $C_t$ and $\textbf{b}_{t-1}$ is given by the vector autoregressive model from \shortciteA{miao2023mis}:$$ \hat{\textbf{b}}_t = C_t \textbf{b}_{t-1} I - \alpha \textbf{b}_{t-1} $$ where the scalar parameter $I$ scales the overall influence of the context matrix, and the negative term $-\alpha \cdot \textbf{b}_{t-1}$ with scalar $\alpha$ ensures stable predictions by preventing unbounded growth. These scalars are fixed to $I=0.5$ and $\alpha=0.9$ for all analyses in this paper.

The key advantage of sequential Bayesian inference is that evidence accumulates in the distribution by recursively forming the prior based on the posterior of the previous timepoint $P(C_{t-1}|\textbf{b}_{t-1}, \textbf{b}_{t-2})$. Typically, the previous posterior undergoes a transition before becoming the prior for inference in the next timepoint. The prior probability of $C_t$ is obtained by summing over all previous posterior probabilities of $C_{t-1}$, each weighted by its probability of transitioning into $C_t$, denoted as $P(C_t|C_{t-1})$. In other words, the prior probability of $C_t$ is the aggregate probability of all previous posterior $C_{t-1}$ transitioning into it: $$P(C_t|b_{t-1}) = \int \underbrace{P(C_t|C_{t-1})}_{\text{Transition}} \,\,\, \cdot \,\, \underbrace{P(C_{t-1}|\textbf{b}_{t-1}, \textbf{b}_{t-2})}_{\text{Posterior at }t-1} \,\,\, dC_{t-1}.$$

This recursive update is intractable since evaluating all possible $C_{t-1}$ (denoted by the integral) is challenging for complex timeseries data in practice. To address this, we implement sequential Bayesian inference with particle filtering. 

\subsection{Particle Filters}
Particle filter implementations represent probability distributions using a set of samples, denoted as particles, that are resampled according to a set of sampling weights. Consider $N$ context matrices as the \textbf{\textit{prior}} at time $t$,  $$\{C_{t}^{(i)}\}_{i=1}^N,$$
each with an associated weight from the previous timepoint $w_{t-1}^{(i)}$. The prior is initialized at $t=2$ (because the context matrix needs two timepoints) as a random sample of 100,000 context matrices drawn from a multivariate normal distribution, $$C_2^{(i)} \sim \mathcal{N}(\textbf{0}_4, I_4)$$ where $\textbf{0}_4$ is a 4-long zero vector and $I_4$ is the $4 \times 4$ identity matrix. The weights are uniform, $w_1^{(i)} = 1 /N = 1/100,000$.

To perform sequential Bayesian inference, follow three steps:
\begin{enumerate}
    \item \textbf{Correction:} Compute the new weight of a particle $w_t^{(i)}$ as the \textbf{\textit{likelihood}} of the particle, normalized across all particles: $$\begin{aligned}
\tilde w_t^{(i)} &= P(\textbf{b}_t \mid C_t^{(i)}, \textbf{b}_{t-1}),\\
w_t^{(i)} &= \frac{\tilde w_t^{(i)}}{\sum_j \tilde w_t^{(j)}}.
\end{aligned}$$
    \item \textbf{Resampling:} Resample particles $C_t$ according to their new weights $w_t$. The result is a \textbf{\textit{posterior}} where high-likelihood particles are replicated and low-likelihood ones are pruned.
    \item \textbf{Prediction:} Perturb each $C_{t}^{(i)}$ by drawing from a transition distribution to obtain the particle for the next timepoint $C_{t+1}^{(i)}$, $$C_{t+1}^{(i)} \sim P(C_{t+1}|C_{t}^{(i)})$$ where, here, the transition distribution is multivariate normal, $$P(C_{t+1}|C_t^{(i)}) = \mathcal{N}(C_{t}^{(i)}, \Sigma _{\text{jitter}})$$  centered at $C_{t}^{(i)}$ with some covariance $\Sigma_{\text{jitter}} = \sigma_{\text{jitter}} I_4$ controlling the extent of the transition. The next timepoint's prior, represented by the samples $\{C_{t+1}^{(i)}\}_{i=1}^{N}$, is known as the \textit{\textbf{temporal prior}}.
\end{enumerate}

The transition distribution generating the temporal prior, specifically its covariance $\sigma_{\text{jitter}}$, is conceptually and computationally important. Conceptually, the covariance dictates the amount of ``jitter" applied to each $C_t^{(i)}$ and therefore controls the extent of ``random exploration" in the particle filter. A small covariance enforces temporal smoothness by keeping particles close to their resampled values (leading to a slow learning rate), while a larger covariance allows particles to spread out (allowing a fast learning rate). The choice of covariance is therefore crucial because it balances fidelity to the data against flexibility, while also mitigating particle uniformity from many successive resamples by injecting diversity.

In the present paper, the appropriate magnitude of transition covariance is not obvious \emph{a priori}. One might consider jointly inferring the context matrix and the covariance within the particle filter, but this is conceptually challenging given a limited amount of observations: the covariance directly governs inference at every step, so attempting to jointly estimate it could destabilize inference---akin to ``determining the best strategy of a game" (inferring context matrix) while simultaneously ``changing the rules of the game" (inferring covariance). Instead, we decouple context matrix inference from covariance estimation by performing grid search, which repeats the entire inference process $k$ times with different covariances. The resulting, more tractable problem is then to select among the $k$ candidate outputs as the ``true" output of the model.

\subsection{Model Outputs}
For each $k$th transition covariance in grid search, the model produces as output two things: (1) a timeseries of most-probable context matrices at every timepoint, $C^* = [C^*_2, \dots, C^*_{t-1},C^*_{t},C^*_{t+1},\dots]$ where $C^*_t = \argmax_{C_t} \big[P(C_t|b_t, b_{t-1})\big]$. And (2) a timeseries of most-probable predicted behaviors, $$\begin{aligned}
\textbf{B}^* &= [\textbf{b}_1, \dots, \textbf{b}^*_{t-1}, \textbf{b}^*_t, \textbf{b}^*_{t+1}, \dots] \\
\textbf{b}^*_t &= C^*\textbf{b}^*_{t-1}I - \alpha \textbf{b}^*_{t-1}
\end{aligned}$$ where $\textbf{b}^*_t = C^*\textbf{b}^*_{t-1}I - \alpha \textbf{b}^*_{t-1}$ is the behavior predicted by the most-probable context matrix $C^*_{t}$ applied to the previous predicted behavior $\textbf{b}^*_{t-1}$, and the first ``predicted" behavior $\textbf{b}^*_1$ is the first observed behavior $\textbf{b}_1$. Recall that $\textbf{b}_t$ is not a single number, but a vector of both agents' behaviors.

\section{Results}

\subsection{Simulation}
To evaluate the accuracy of the sequential Bayesian model (Criterion 1), we generated a set of ground-truth context matrices from \shortciteA{miao2023mis}. We created 81 matrices by permuting four entries, where every entry took one of three values (-1, 0, 1). For each matrix, we defined 5 noise levels. At each noise level, the model produced two timeseries (one per agent, 500 points long). See Method for details.

\begin{figure}
\centering
\includegraphics[width=0.8\textwidth]{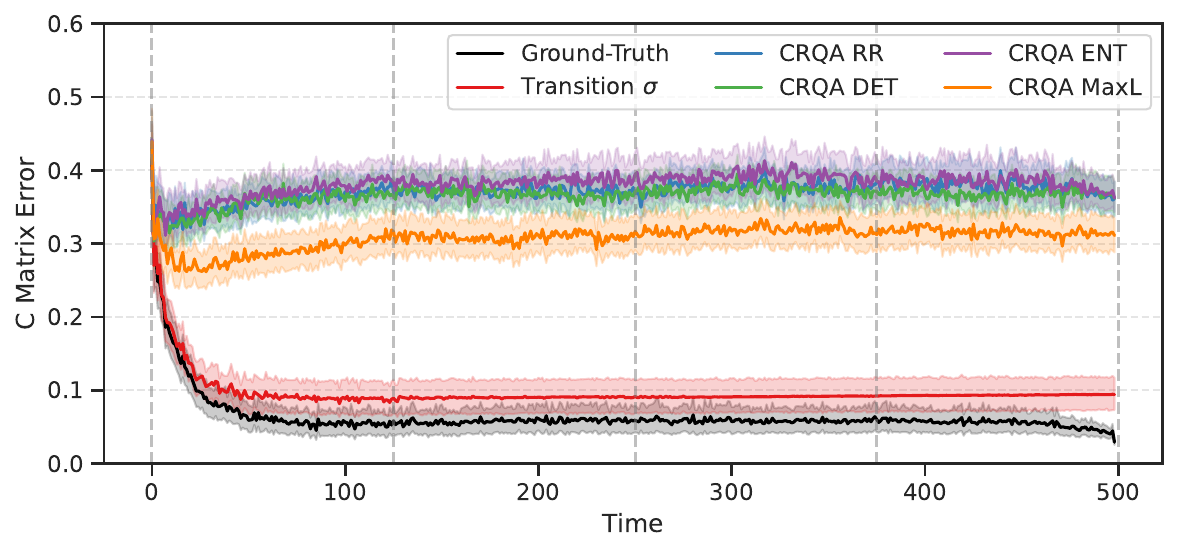}
\caption{Context matrix error over time for different methods of selecting transition covariances. Ideally, the model consistently chooses the covariance whose inferred context matrices are closest to ground-truth (black). This is the model's ceiling performance. When ground-truth is unknown, but we assume only one context matrix governs data, error approaches ceiling (red). Without knowing ground-truth and assuming constancy, we select covariances with highest similarity to observed behavior along various measures of synchrony (colored lines). Vertical lines delimit quartiles. Error bands are 95\% CI.}
\label{fig:sim_figureA}
\end{figure}

\begin{figure}[!b]
\centering
\includegraphics[width=0.8\textwidth]{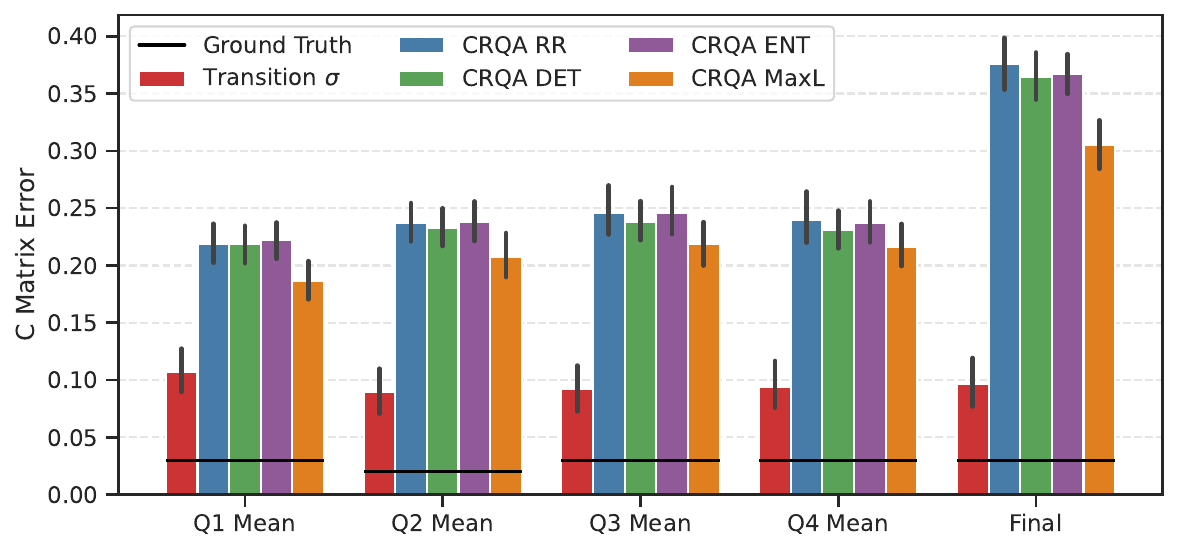}
\caption{Context matrix error using quartile means of context matrices or the final context matrix for different covariance selection methods. Black lines indicate use of context matrices from covariances whose outputs have minimal error. Colored bars are from outputs selected by synchrony measures. Error bars are 95\% CI. Note that overall errors are lower than in Figure~\ref{fig:sim_figureA} due to different averaging schemes.}
\label{fig:sim_figureB}
\end{figure}

Given ground-truth, we can approximate the best possible performance of the model by identifying the appropriate transition covariance value from grid search. In the \textit{ideal case}, the model consistently selects transition covariances whose inferred context matrices have the smallest errors relative to ground-truth. We define error as the mean absolute difference between the four cells of the inferred and ground-truth context matrices. Errors of 0.5 or greater can be considered high because they yield qualitatively different interpersonal dynamics \shortcite{miao2023mis}. A key question is how to compute error given the model produces a \textit{timeseries} of context matrices, not a single point estimate. Figure~\ref{fig:sim_figureA} plots error over time across all simulations. The black line represents selecting covariances whose inferred context matrices have minimal error. As shown, error falls sharply within the first quartile (0-125) and remains low onward. Here, we use either the final inferred matrix, an estimate based on all available information in the timeseries, or an aggregate of matrices over quartiles of the timeseries. When error is computed from the mean of matrices within the second quartile, it is lowest, $M = 0.02, sd=0.06$. Using the final matrix produces similarly low error, $M = 0.03, sd=0.04$ (Figure~\ref{fig:sim_figureB}, black horizontal lines). 

The question arises how to select a transition covariance without ground-truth knowledge, as is the case with human data. As an intermediate step, we consider a slightly more realistic \textit{toy case} in which the model does not know ground-truth but assumes that an observed timeseries is governed by a single, constant context matrix---unlike the human setting where interpersonal dynamics may shift within a task. Under this constancy assumption, there is no problem of selecting a covariance value because the optimal strategy is to fix it to a low value. The reason is that a large covariance or ``jitter" introduces unnecessary variability to a posterior with high probability mass on what may be the correct context matrix. By contrast, a low covariance reduces exploration of context matrices, which is a more appropriate inductive bias when assuming constancy. Indeed, using a low-covariance selection scheme, error approaches the ideal case (Figure~\ref{fig:sim_figureA}). Specifically, error is $M = 0.09$ when using second, third, and fourth quartile means (Figure~\ref{fig:sim_figureB}) compared to $M\geq 0.22$ when using the highest covariance values.

\begin{figure}
\centering
\includegraphics[width=\textwidth]{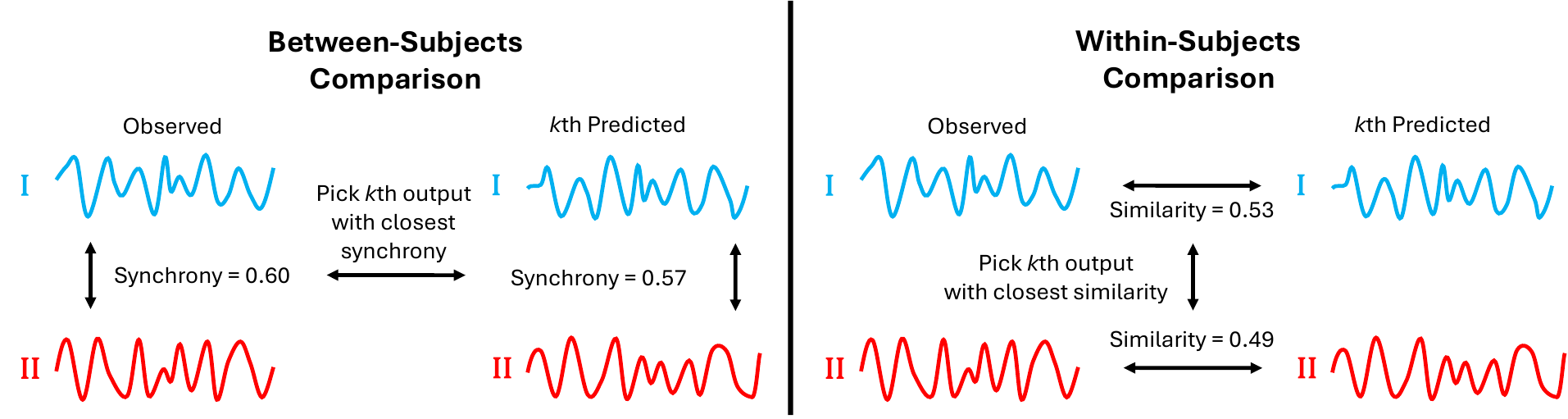}
\caption{Between-subjects and within-subjects style of selecting the \textit{k}th covariance value without knowledge of ground-truth context matrices. The context matrix implicitly encodes synchrony as a pattern of cross- and self-influence. Using between-subjects comparison, the \textit{k}th predicted behavioral timeseries with minimal context matrix error should reflect similar synchrony as the observed timeseries. Within-subjects indirectly preserves synchrony by selecting \textit{k} that preserves agents' timeseries to observed.}
\label{fig:sim_figureC}
\end{figure}

What if we cannot assume that interpersonal dynamics remain constant? In the \textit{realistic case}, we test different covariance values via grid search and must choose among them without ground-truth. A natural strategy is to select a value so that its predicted behavioral timeseries best match the observed behavioral timeseries. Since predicted timeseries are generated from the inferred context matrices, their similarity to observed behavior may implicitly track context matrix error, such that closer matches imply more accurate estimates. Recall that, for a given covariance value, four timeseries are relevant: the predicted and the observed for each agent. These timeseries can be compared in two ways. Between-subjects comparison computes a relation feature between agents (e.g., synchrony) separately for predicted and observed, and defines similarity as a close match between relations (Figure~\ref{fig:sim_figureC} left). Within-subjects comparison computes a similarity-like score between predicted and observed, separately by agent, and defines similarity as the average of scores (Figure~\ref{fig:sim_figureC} right). Conceptually, between-subjects similarity emphasizes whether the relation between agents’ timeseries is preserved, whereas within-subjects similarity evaluates whether each agent’s timeseries is preserved.

In what follows, we report results using an existing operationalization of synchrony as the metric. The reason is that synchrony is explicitly encoded in the context matrix as a pattern of cross- and self-influence, so when a predicted timeseries exhibits a level of synchrony between agents similar to the synchrony observed between agents, this suggests a good fit in interpersonal dynamics. Here, synchrony is operationalized with CRQA \shortcite{coco2014cross}. CRQA produces four metrics for a pair of timeseries: recurrence rate (RR), determinism (DET), entropy (ENT), and max length (MaxL). These four metrics quantify different aspects of synchrony. RR reflects how often two agents revisit the same temporal state (i.e., shared behavioral motif). DET measures the extent to which these revisits form sustained sequences of motifs. ENT captures the diversity of these sequences' lengths. MaxL measures the length of the longest sequence. 

We report results based on between-subjects similarity of CRQA in the main text and describe within-subjects CRQA results in the Supplementary Information. Between-subjects is preferred because it directly preserves a \textit{relation} between agents, in this case synchrony, and interpersonal dynamics are relations of a broader kind. For between-subjects similarity, we select the covariance that leads to behavior output with the smallest difference in a CRQA metric relative to observed. Figures~\ref{fig:sim_figureA} and~\ref{fig:sim_figureB} show that MaxL produces lowest error across methods of selecting the context matrix. The lowest MaxL error uses context matrix means of the first quartile, $M = 0.19, sd=0.18$, and the highest error uses the final matrix, $M=0.31, sd=0.22$. For within-subjects similarity, the largest average of a CRQA metric between an agent's predicted and observed timeseries is chosen. Highly similar errors emerge, $M \geq 0.18$ (Supplementary Information).

In sum, we provide evidence that sequential Bayesian inference can infer context matrices with sufficient accuracy even with noise (Criterion 1). Error was nearly perfect with ground-truth knowledge, $M = 0.02$, remained low without ground-truth under a constant-matrix assumption, $M = 0.09$, and was well-within the range of the correct dynamic without knowing ground-truth or assuming constancy, $M \approx 0.18 \leq 0.5$. In the Supplementary Information, we further explored usage of non-CRQA metrics (e.g., correlation). All, however, yielded comparable errors. We take this to suggest that, while accuracy does not reach ceiling performance in a realistic situation, the model is not brittle: its success does not depend on CRQA, and different reasonable operationalizations of similarity between observed and predicted behavior converge on the same results. Such robustness is valuable for human data, where ground-truth is unknown and no single metric can be assumed correct.

\subsection{Human Experiment}

\begin{figure}[!b]
\centering
\begin{subfigure}{\textwidth}
    \centering
    \includegraphics[width=\textwidth]{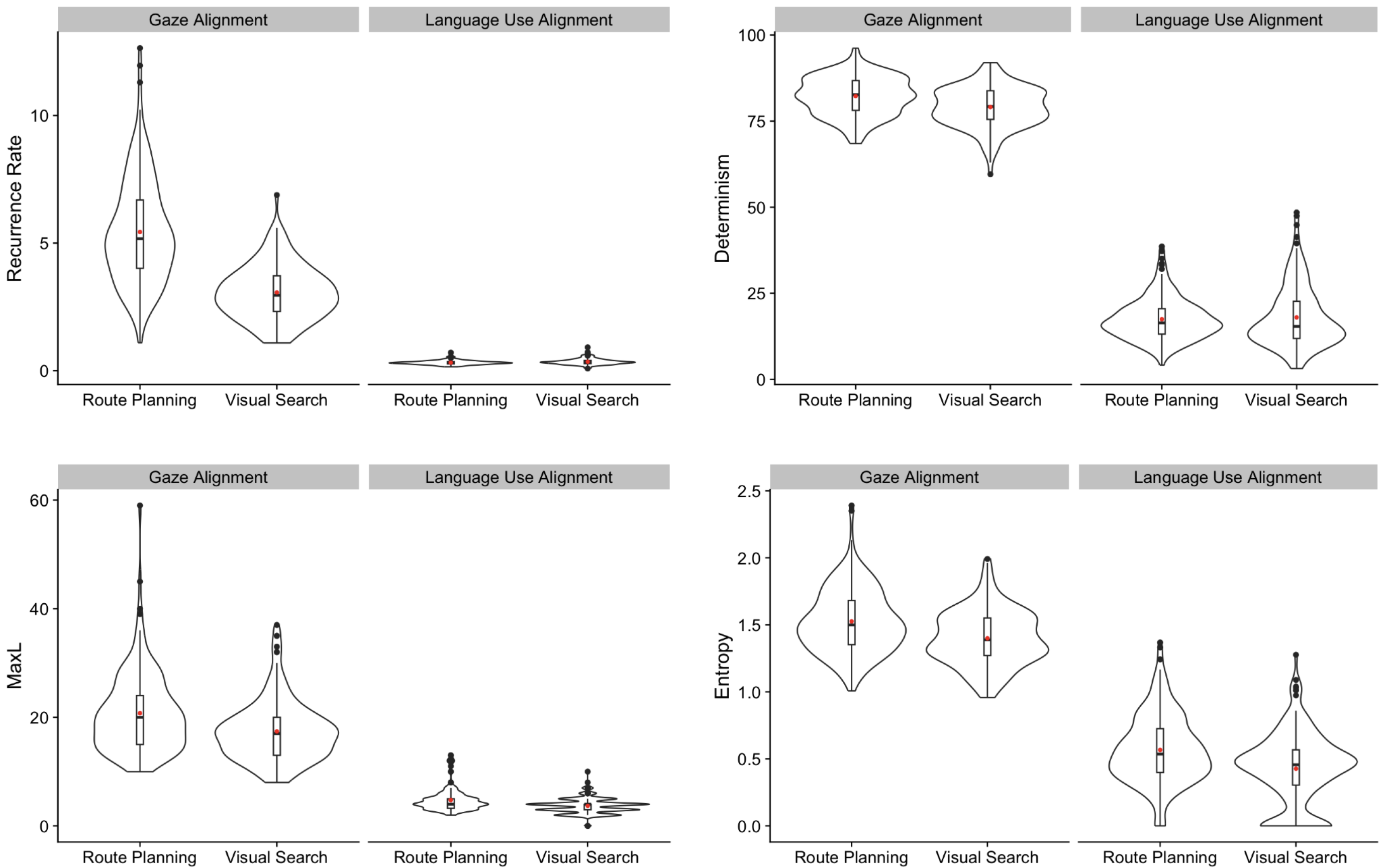}
    \caption{}
\end{subfigure}
\hfill
\begin{subfigure}{0.45\textwidth}
    \centering
    \includegraphics[width=\textwidth]{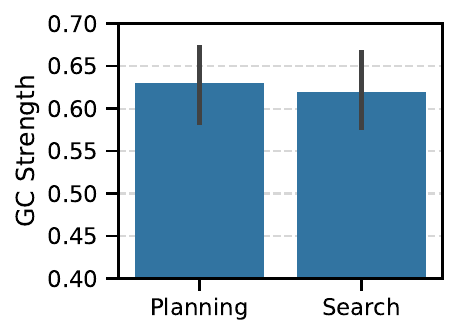}
    \caption{}
\end{subfigure}
\caption{Task-based differences of observed data, independent of our model. (a) CRQA values are higher in planning than search. Figure from Galati et al. (accepted). (b) GC strength does not differ across tasks. Error bars are 95\% CIs.}
\label{fig:hum_figureAB}
\end{figure}

A psychologically meaningful representation of interpersonal dynamics is not only recoverable under noise (Criterion 1) but should prove useful by varying systematically with task-based differences in interpersonal dynamics (Criterion 2). To make this claim, we ran a human experiment in which dyads completed two tasks: route planning and visual search \shortcite{galatietal}. In route planning, dyads identified the most efficient path between two subway stations; in visual search, they counted stations with a specified feature (e.g., the number of stations that start with a particular letter or contain a specific word). Each planning and search trial was coded with a binary accuracy (1 or 0). In a separate paper, we showed that planning elicits greater synchrony in speech and eye-gaze, as quantified by CRQA metrics, than visual search \shortcite{galatietal}. Figure~\ref{fig:hum_figureAB} shows results from that paper. Here, we also find that planning yields similar leader-follower dynamics as search in eye-gaze, as measured by Granger causality, $\beta=0.025, t=1.13,p=0.27$ (Method). If context matrices vary systematically by task, these patterns should qualitatively appear in inferred context matrices from human data.

We applied the sequential Bayesian model to each participant pair's eyetracking data from these tasks (Method). To compare the inferred context matrices across tasks, we needed to define features of comparison. A context matrix for two individuals contains four values (two self-influences and two cross-influences) but taken together these numbers are not readily interpretable at the group level. We therefore constructed summary features (from the matrix) that quantify synchrony and leader-follower dynamics and are roughly analogous to existing measures like CRQA and Granger causality. To quantify synchrony, we constructed \textit{relative influence}, the proportion of agents' behaviors attributable to each other versus themselves, $$\text{Relative Influence} = \frac{|c_{12}| + |c_{21}|}{\sum_{i,j} |c_{ij}|}$$ where indices $i$ and $j$ refer to a row and column of the context matrix. High relative influence (near 1) indicates strong coordination, with each agent largely driven by their partner’s previous behavior; low values (near 0) indicate independence. Relative influence is broadly analogous to CRQA, but conceptually different. To quantify leader-follower dynamics, we defined two summary features. First, \textit{leader strength} captures the relative strength of the agent with higher cross-influence, $$\text{Leader Strength} = \frac{\big||c_{12}| - |c_{21}|\big|}{|c_{12}| + |c_{21}|}$$ where high leader strength indicates a strong leader and low values indicate an egalitarian dynamic. Leader strength is conceptually and mathematically similar to GC strength in Granger causality. Second, \textit{leader switch rate} is the frequency of leader-follower switches, defined as the proportion of times the sign of the cross-influence difference $|c_{12}| - |c_{21}|$ flipped within a trial (where the denominator is the number of consecutive non-zero timepoints). To our knowledge, leader switch rate has no analog in existing methods because it leverages the time-resolved nature of the context matrix rather than collapsing behavior into a single aggregate score, as in non-time-resolved measures like CRQA and Granger causality.

To compare these summary features across tasks, relative influence and leader strength were computed at each timepoint, then summarized as quartile means, overall trial means, or final values; leader switch rate was computed for quartiles and whole trials. Importantly, all analyses used mixed-effects regressions with a summary feature as the outcome and various deliberately chosen random effects, including for covariance selection metric (four CRQA metrics from simulation), trial index, index of quartile or final (coded as a ``quartile," unless leader switch rate is the outcome), and participant pair, which had a random intercept for task type since tasks were within-subjects. The random effects of covariance selection metric and index of quartile/final were included to test whether an effect holds across multiple metrics and quartiles/final by accounting for differences between metrics and between quartiles/final. The random effect of trial index ensures that covariances re-selected across metrics are not double-counted as independent data points, preventing artificial inflation of statistical power. This consolidated mixed-effects regression approach avoids inflating Type I error from otherwise testing every combination of task type, covariance selection metric, and quartile/final.

\begin{figure}
\centering
\includegraphics[width=0.7\textwidth]{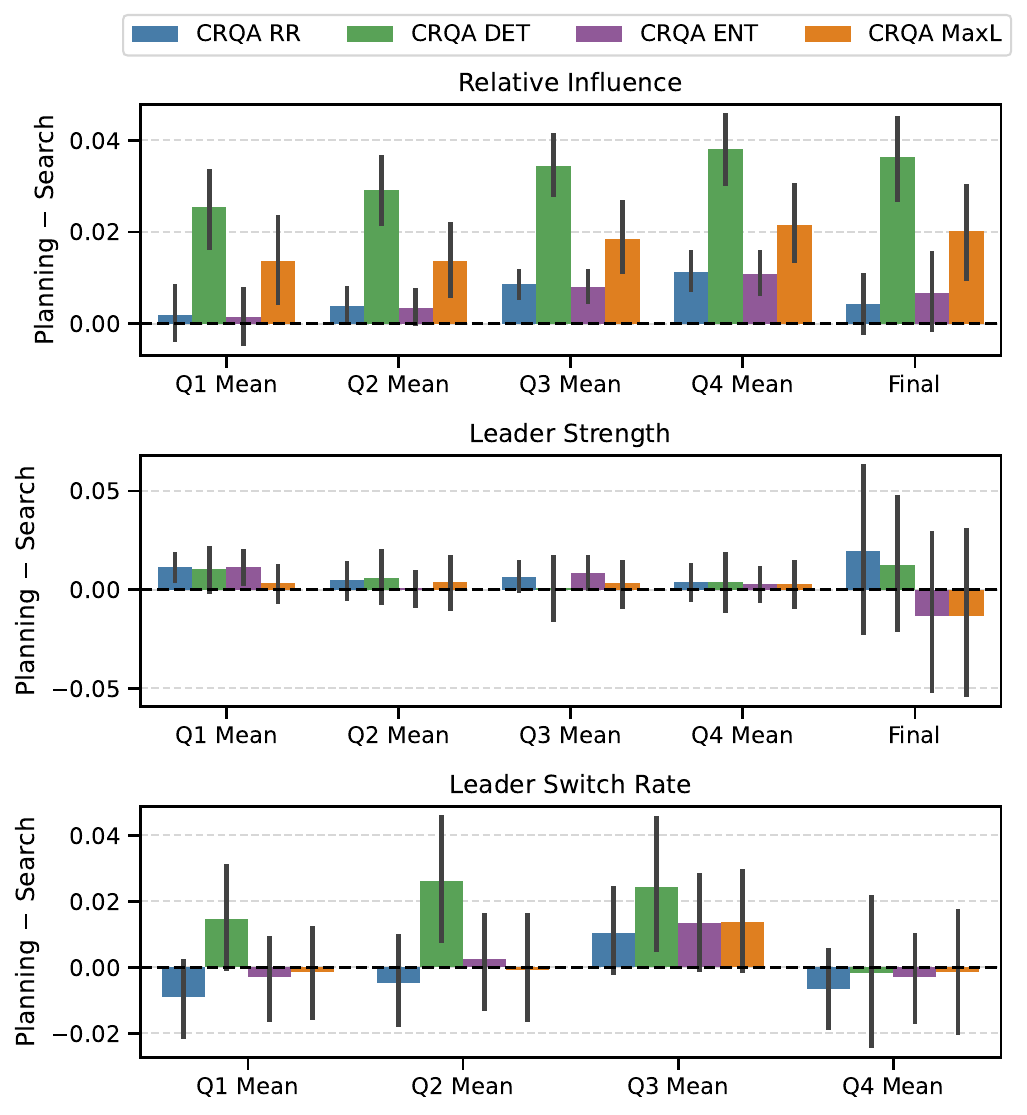}
\caption{Differences in relative influence, leader strength, and leader switch rate between route planning and visual search trials when selecting transition covariances using CRQA metrics (between-subjects comparison). Positive values mean more quantity is present in planning trials; negative means more in search; and zero means equal. Error bars are 95\% CIs.}
\label{fig:hum_figureC}
\end{figure}

Just as CRQA and Granger causality showed higher synchrony in planning compared to search and similar leader-follower dynamics in both, the inferred context matrices reflect the same pattern: higher relative influence in planning and similar leader strength and leader switch rates in both (Figure~\ref{fig:hum_figureC}). Three mixed-effects models, each with a different summary feature as the outcome, confirmed this (relative influence: $\beta=0.016, t=6.47, p<0.0001$; leader strength: $\beta=0.0054, t=1.74, p=0.09$; leader switch rate: $\beta=0.0045,t=1.03,p=0.31$). Interestingly, we also found that relative influence, but not leader strength or leader switch rate, increased over time in a trial (relative influence: $\beta=0.0035, t=6.30, p<0.0001$; leader strength: $\beta=-0.0017, t=1.77, p=0.08$; leader switch rate: $\beta=-0.000058,t=0.037,p=0.97$). This was tested with mixed-effects models with fixed effects of task type, quartile index (removing final), and their interaction, and observing the significance of the interaction. 

\begin{figure}[!b]
\centering
\includegraphics[width=0.7\textwidth]{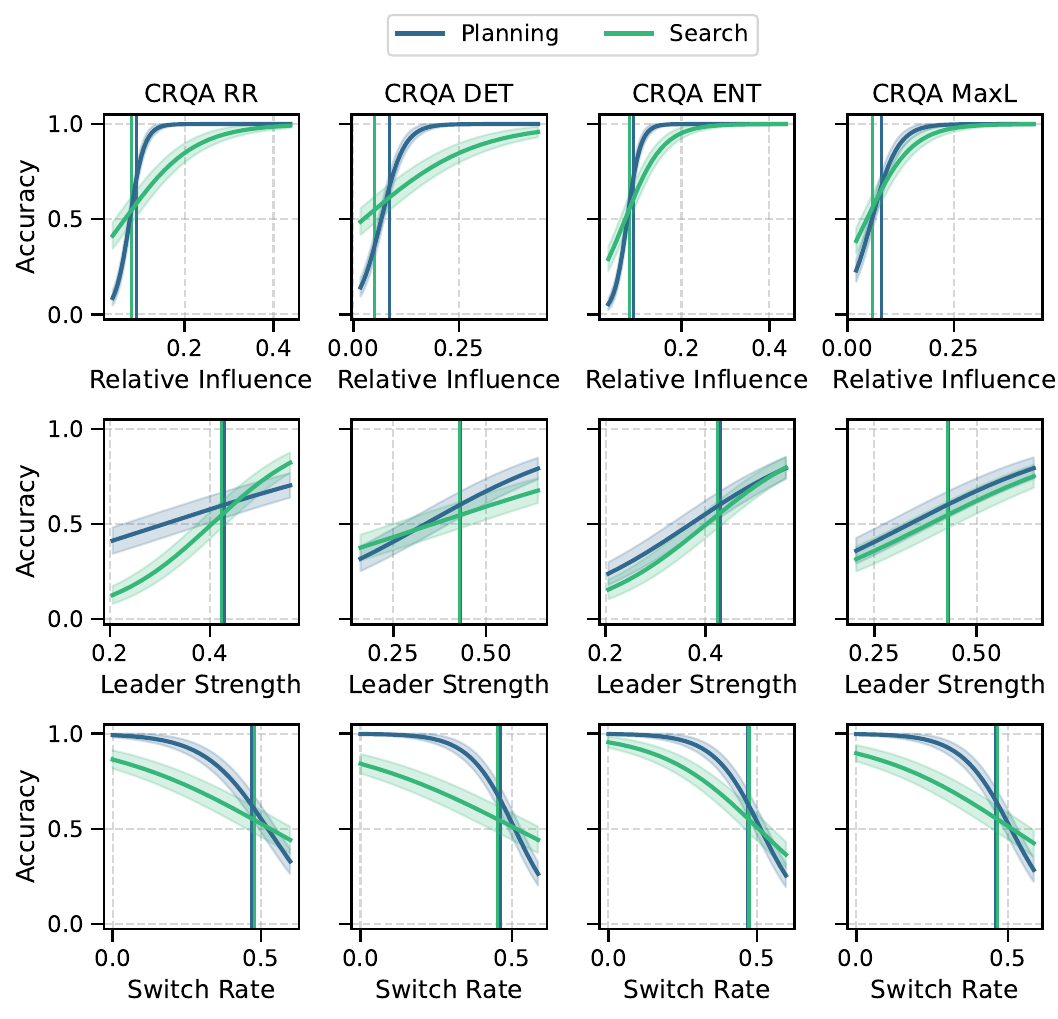}
\caption{Logistic regressions of trial accuracy against trial means (not quartiles) of relative influence, leader strength, and leader switch rate, separately for planning and search tasks and for each selection metric (CRQA). Vertical lines show average overall means on correct trials. Error bands are 95\% CIs.}
\label{fig:hum_figureD}
\end{figure}

If context matrices can capture qualitative patterns of task-based variations in interpersonal dynamics, could they go beyond and predict final decision outcomes, like a trial's accuracy (``correct" or ``incorrect"), in reasonable directions (e.g., more synchrony increases accuracy for planning)? We previously showed that greater CRQA synchrony in planning \textit{and} search tasks predicts higher accuracy \shortcite{galatietal}. Figure~\ref{fig:hum_figureD} suggests that context matrices capture this pattern as well: higher relative influence, higher leader strength, and fewer leader switches predict higher accuracy in both tasks. To test this, we used three logistic mixed-effects models with accuracy as the outcome, and task type, the summary feature (as overall trial means) derived from the context matrix, and their interaction as fixed effects, removing the random effect for quartile/final. 

The models partially supported these trends. First, the interaction between task type and leader strength is not significant, suggesting that task types do not differ in their relationship between leader strength and accuracy, $\beta=6.81,z=1.52,p=0.13$. Collapsing over tasks, leader strength predicts accuracy, $\beta=7.74,z=2.10,p=0.036$. Second, the interactions are significant for relative influence and leader switch rate, $p\leq 0.015$. Follow-up simple effects reveal that accuracy of planning trials is positively predicted by relative influence, $z=3.51,  p<0.001$, and negatively predicted by leader switch rate, $z=-4.31 , p<0.0001$. However, accuracy of search trials is not predicted by relative influence, $z=-0.73 ,p=0.47$, or leader switch rate, $z=-1.73,p= 0.089$. This null effect of relative influence on search accuracy is consistent with our earlier claim that search often elicits a divide-and-conquer strategy: participants split up and search different areas \shortcite{galatietal}. Here, both synchronous search and divide-and-conquer can be effective, so synchrony is not a reliable predictor of accuracy.

Given that planning yields higher relative influence than search, and that relative influence predicts higher accuracy in planning, one might expect higher accuracy in planning overall. However, accuracy did not differ between tasks, $z = 1.1, p = 0.26$. This discrepancy is explained by the fact that the raw difference in relative influence was too small to elicit a statistically detectable accuracy difference. In Figure~\ref{fig:hum_figureD}, the vertical lines in the first four plots represent relative influence for correct planning and search trials, and their intersections with logistic curves indicate predicted accuracy. The difference in these predicted accuracies ranges from $18.1\%  - 22.1\%$, small enough to remain statistically undetected. 

Next, we examined how relative influence, leader strength, and leader switch rate derived from the estimated context matrices relate to direct behavioral measures of synchrony like CRQA and of directional influence like Granger causality. Recall that CRQA broadly quantifies recurrence or repetition in shared temporal states, whereas relative influence captures behavioral dependence. High relative influence does not guarantee recurrence---for example, freestyle dancers (low DET) that rarely return to the same moves (low RR) with little variety in move sequence durations (low ENT) and short-lived longest move sequence (low MaxL), but nevertheless respond to each other's moves (high relative influence). Therefore, although both relative influence and CRQA metrics are higher in planning than search at the group level, they need not positively covary at the trial level. In contrast, leader strength may overlap with Granger causality at the trial level because leader strength is conceptually similar to GC strength: both are relative imbalances in directional influence magnitudes.

Analyses suggest that relative influence is distinct from CRQA values and that leader strength partially overlaps with Granger causality. For relative influence, we fit four mixed-effects models predicting trial-mean relative influence from each CRQA metric, with random effects as before (removing quartile). All CRQA metrics negatively predicted relative influence (RR: $\beta=-0.082,  t=-3.07, p<0.01$; DET: $\beta=-0.094, t=-13.96, p<0.0001$; ENT: $\beta=-0.0065, t=-15.02, p<0.0001$; MaxL: $\beta=-0.00084, t= -8.88,p  < 0.0001$). This confirms that CRQA and relative influence are distinct conceptions of synchrony. For leader strength, we first fit a mixed-effects model predicting trial-mean leader strength from GC strength, and found that GC strength did not predict leader strength, $\beta=0.0072, t= 1.49, p =0.14$. However, setting fixed effects to GC strength, task type, and their interaction revealed a significant interaction, $\beta=-0.026, t=-2.67, p\leq0.01$. Simple effects reveal that GC strength predicts leader strength for planning trials, $t=2.88,p<0.01$, but not search, $t=-0.89, p=0.38$.

Together, these findings provide converging evidence that inferred context matrices capture human interpersonal dynamics. First, inferred context matrices have discriminative power: summary features of the context matrix replicate task-based variations (or lack thereof) in the observed data (as reported here and by \shortciteNP{galatietal}). Second, the matrices have predictive utility: those features also explain trial-level outcomes like accuracy in theoretically reasonable directions (e.g., planning accuracy increases with more synchrony; \shortciteNP{wallot2016beyond, gurtner2007getting}). Third, the findings form consistent narratives: the fixed effect(s) of interest remains (non-)significant when covariance selection metrics are set to a random effect, indicating that a result holds across different metrics. While alternative interpretations regarding the accuracy findings remain plausible (e.g., higher leader strength could justifiably reduce planning accuracy), the overall evidence appears to favor the context matrix as a useful representation more than not.

\section{Discussion}

Teams succeed or fail based on their interpersonal dynamics, the ways members coordinate, influence, and adapt to one another \shortcite{janis1972victims, wallot2016beyond, brennan2008coordinating, gurtner2007getting}. Task demands shape interpersonal dynamics \shortcite{coco2018performance, galatietal}, yet a unified psychologically meaningful measure of it is missing. Prior measures typically capture single dimensions of interpersonal dynamics, such as synchrony (e.g., CRQA, correlation) or direction of influence (e.g., Granger causality, transfer entropy), but rarely at moment-to-moment resolution \textit{and} in a form that easily scales to larger teams. Here, we advance the \textit{context matrix} \shortcite{miao2023mis} as a psychologically meaningful representation of interpersonal dynamics at timepoint-specific resolution for any group size, from which we derive interpretable values of synchrony (relative influence) and direction of influence (leader strength, leader switch rate). We evaluated this claim against two criteria: (1) context matrices should be inferrable with sufficiently high accuracy, and (2) inferred context matrices should vary systematically across tasks that elicit different interpersonal dynamics. 

\subsection{Accurately Inferring the Context Matrix}

To provide evidence for the first criterion, we developed a sequential Bayesian model to infer context matrices from noisy simulated timeseries. In sequential Bayesian inference, a probability distribution over possible context matrices becomes increasingly refined as evidence accumulates over time. We implement this as a particle filter, a discrete sample of context matrices that is resampled according to a corresponding set of weights. When the data strongly support one context matrix, that context matrix will correctly dominate the set. But if the context matrix underlying the data changes, this overly homogenous set will struggle to change. To increase flexibility, particle filters randomly alter the samples at every timepoint according to the covariance of a transition distribution. Since determining the appropriate magnitude of this covariance is challenging \textit{a priori}, we re-run inference using a wide range of possible values (through grid search). The problem then becomes determining the optimal value resulting in minimal context matrix error. 

We reasoned that an optimal covariance value among those tested should produce predicted data (predicted from inferred context matrices) that best exhibit the \textit{relations} between agents in the observed data. The reason is that the context matrix fundamentally characterizes pairwise relations between agents; predicted data from inferred context matrices that preserves these relations may therefore track context matrix error. We found that covariance values producing predicted data that preserves the observed synchrony between agents produced accuracy that, though imperfect, was within the range of a correct context matrix---enough, we believe, to claim the model is sufficiently accurate to apply to human data. There is, nevertheless, room for improvement: the model's ceiling performance is near perfect, and performance is high when assuming that the context matrix underlying the data is constant. Additional attempts at reducing error resulted in similar performance (Supplementary Information), suggesting that future work should focus on exploring alternative model refinements rather than incremental adjustments.

\subsection{The Context Matrix as a Psychologically Useful Representation}

To provide evidence for the second criterion, we applied the model to human eyetracking data from dyads completing route planning and visual search tasks. We found that, compared to search trials, planning trials produce higher synchrony and similar leader-follower dynamics in dyads, measured through CRQA and Granger causality (reported here and by \shortciteNP{galatietal}). Inferred context matrices from these data qualitatively mirrored these patterns: they displayed higher \textit{relative influence} in planning trials (a measure of synchrony derived from the context matrix) and similar \textit{leader strength} and \textit{leader switch rate} between tasks (also derived from the matrix). 

Beyond reproducing these patterns, the context matrices also predicted trial-level accuracy in theoretically meaningful ways. Higher relative influence predicted higher accuracy in planning trials, consistent with prior work showing that increased team coordination benefits performance \shortcite{wallot2016beyond, gurtner2007getting}. Higher leader strength and fewer leader switches also predicted higher accuracy in planning, consistent with the idea that stronger leaders can reduce negotiation overhead by centralizing decision-making and fewer leader-follower switches avoid disruptions to shared strategy. In contrast, search trial accuracy was not predicted by relative influence or leader switches, and was only indirectly implied to depend on leader strength given the absence of an interaction with task type. Although the latter result does not replicate our earlier patterns with CRQA, it too aligns with our larger psychological distinctions between planning and search. Accuracy in visual search (locating a target among distractors) can be achieved through more independent roles without stable/strong leadership \shortcite{galatietal}. For example, participants often adopted a ``divide and conquer'' strategy, in which they searched different areas of the map for landmarks independently and then pooled their counts to obtain a final answer. This suggests that, while planning accuracy hinges on coordinated leadership, search accuracy can emerge from parallel, relatively independent contributions.

Furthermore, relative influence and leader strength of the context matrix appear to relate to existing measures (CRQA and Granger causality) in expected ways. Although CRQA and relative influence are both operationalizations of synchrony, they are distinct conceptions of it: CRQA measures recurrence or repetition in shared temporal states, whereas relative influence measures agents' overall mutual influence on each other's behavior. In line with this, we find that these measures negatively covary. Likewise, Granger causality and leader strength appear conceptually similar. We define GC strength as the relative imbalance in the effect size of predicting one agent's behavior from the other's (Method), and leader strength as the relative imbalance in cross-influence terms (Results). We find that GC strength positively covaries with leader strength in planning trials but not in search, perhaps owing to the fact that these measures are not identical. 

One interpretation of these findings is that participants are dynamically assembling what some have referred to as ``task configurations,'' a set of behavior patterns that operate to stabilize some level of performance \shortcite{riley2011interpersonal}. Our sequential Bayesian model frames this dynamic as a kind of sampling over possible \textit{relative} behavioral patterns expressed in context matrices. Context matrices may in these cases infer the underlying configuration that participants are approximating, even if it does not covary with task performance itself---the configuration stabilizes some level of desired or necessary performance.


\subsection{Beyond the Context Matrix}

Prior measures of interpersonal dynamics are typically treated as approximations of one dimension of interaction, not as representations of the whole dynamics themselves. Although we propose the context matrix is a unified, psychologically useful representation, how well the estimated matrix could truly reduce error remains an open question. The context matrix currently assumes agent-agent causal influence is the basic unit of interaction and models behavior as a linear function of past states. These are simplifications, and future work should test variations, for example by aggregating influence terms to capture high-level dynamics in larger teams or using more complex linear or non-linear functions for behavior.

A key strength of the context matrix lies in its pairwise format: each cell encodes interactions between agents, including self-pairs, making all pairwise dependencies explicit and allowing asymmetrical relationships (e.g., agent 1’s effect on agent 2 $\neq$ agent 2’s effect on agent 1). While the context matrix currently encodes influence, the same structure could represent other values, such as trust, deference, or approval. A more complete representation may eventually require moving beyond the current formulation, but the pairwise nature of the matrix may be a fundamental building block for successful frameworks of interpersonal dynamics.

Overall, sequential Bayesian inference is well-suited for estimating context matrices. It ensures that accurate matrices are discovered by evaluating the likelihood of many possible matrices and accumulating evidence in the prior over time. However, by assuming a first-order Markov property, sequential Bayesian inference suffers from the limitation that the probability of a context matrix depends only on the immediately preceding timepoint. Human teams, by contrast, rely on memory of longer histories to selectively weigh past events by salience or relevance. The hidden Markov formulation treats all prior states as equally informative, which fails to capture long-horizon dependencies. A more ideal inference model would allow non-uniform integration of past information.

\section{Method}

\subsection{Simulation}
Each simulation began by randomly sampling initial behaviors for two individuals, with values uniformly drawn between 0 and 1. At each subsequent timepoint, noise was independently added to each individual's behavior, sampled from a uniform distribution $[-a, a]$ where $a\in[0.05, 0.10, 0.15, 0.20, 0.25]$. Thus, we created a total of $81 \times 5 = 405$ simulated datasets. For each of these datasets, we repeated the inference procedure with $k = 51$ transition covariances. The square root of these covariances (i.e., standard deviations) ranged incrementally from 0.001 to 0.1 in steps of 0.002. We used the vector autoregressive model from \shortciteA{miao2023mis} to generate a timeseries: $$\textbf{b}_{t} = C\textbf{b}_{t-1}I - \alpha\textbf{b}_{t-1}.$$where $\alpha$ and $I$ were fixed at 0.1 and 0.5, respectively. These values were also used by the sequential Bayesian model.

To compute CRQA, we used the following hyperparameter values. Time-delay embeddings had dimensions $m = 3$ and delay $\tau = 2$. Distances were thresholded at $\epsilon = 0.84$ in z-score units. DET was computed as the proportion of recurrent points forming diagonal lines of length $\geq l_{min} = 2$.



\subsection{Human Experiment}
A total of 80 participants (40 pairs) from the UNC Charlotte community completed the experiment. Participants included undergraduate students, graduate students, and staff. Participants were recruited as strangers; in 4 instances, pairs happened to know each other in advance. All experimental protocols were approved by the IRB committee at UNC Charlotte. All methods were carried out in accordance with relevant guidelines and regulations. All participants provided informed consent.

Participants performed five route planning and five visual search trials, counterbalanced in order. All the subway maps across trials were different (10 unique maps). Their presentation was controlled such that there were 2 sets of 5 maps, matched in difficulty, which were paired with the route planning and visual search tasks. The pairing of task and map set was counterbalanced across pairs. Within each set, the difficulty of the map increased with each trial. Eyetracking data (pupil dilation and x- and y-gaze coordinates), were collected as participants viewed the same map displayed on two separate computer screens, one for each participant. Participants were seated at adjacent workstations, positioned 93 cm from their respective monitors (display size approximately 18 degrees of visual angle; resolution: 1920 × 1080 pixels). Eye movements were recorded using a desktop-mounted SR EyeLink 1000 Plus eyetracker, operating at a 1000 Hz sampling rate. For full details about the experiment, see \shortcite{galatietal}. Given the variability in temporal sampling of eyetracking data, we standardized the temporal resolution across all trials by averaging the behavioral samples within uniform time increments of 272 ms, the median sampling duration across the entire dataset. When pairs had differing numbers of samples within a trial, we truncated the longer timeseries to match the shorter one. 

In the Results, we reported that planning and search trials elicit different leader-follower dynamics. To test this, we applied Granger causality to the observed data to test whether past activity in one agent improved the prediction of behavior in the other agent, and vice versa. To apply Granger causality, we first z-scored the timeseries. We then fit bivariate vector autoregressive (VAR) models with an intercept and possible lag orders 1 to 12. The optimal lag order was chosen by minimizing Bayesian Information Criterion (BIC). Using the optimal lag order, two likelihood-ratio (LR) tests were performed, one for agent 1 predicting agent 2, and another for agent 2 predicting agent 1. The null hypotheses of these tests are that the agent's past values do not improve prediction of the current values of the other agent, conditioned on the other agent's past values. The LR test produces an LR statistic. The LR statistic can be considered the effect size of Granger causality, denoted as GC. When GC values were unstable (i.e., infinite), we repeated the analysis on first differences. Each observed timeseries' behavioral channel (x-pos, y-pos, pupil size) produced two GC scores, one per direction. For each channel, we computed a ``GC strength" score as follows: $$\text{GC Strength} = \frac{|GC_{1\rightarrow2} - GC_{2\rightarrow1}|}{GC_{1\rightarrow2} + GC_{2\rightarrow1}}$$ and then averaged across channels to get an overall GC strength. Note that the formula of GC strength is similar to our formula of leader strength in the context matrix. We tested whether GC strength differed across tasks with a mixed-effects model predicting GC asymmetry from task type, with a random effect of participant pair, which had a random intercept of task type, as it is a within-subjects variable. 

To apply sequential Bayesian inference on human data, the model performed inference using all five trials of a task at once, as trials were blocked by task. In addition, we had to account for three behavioral channels per participant (pupil dilation, x-position, and y-position). First, gaze positions (x and y coordinates) were converted into a discrete spatial grid, dividing the computer screen into a coarse-resolution grid of 10 horizontal (x-axis) by 5 vertical (y-axis) boxes. Second, all three behavioral measures for each participant were standardized within each trial, ensuring comparability across channels. Third, the likelihood of a candidate context matrix in the particle filter was computed separately for each behavioral channel. The joint likelihood for a given context matrix was then calculated by multiplying the individual likelihoods across all three behavioral channels. Since the model simultaneously infers a single context matrix using all three behavioral channels at each timestep, it also produces corresponding predicted behavioral timeseries for each channel individually. For covariance selection, CRQA metrics of a predicted timeseries were computed by taking the score for every channel (pupil dilation, x-position, y-position) and then averaging across channels.

\section{Acknowledgments}
This material is based upon work supported by the National Science Foundation under Grant No. 2120932. 

\section{Author Contributions}
A.J.L., G.M., R.D., and H.L. conceived of the project. H.L. and A.J.L. developed the sequential Bayesian model. A.J.L planned and conducted all simulation and human data analyses. A.J.L. wrote the initial manuscript. A.G. provided the human data. A.G., R.D., G.M., H.L., and A.J.L. revised the manuscript. A.G., R.D., and H.L. supervised the project.

\section{Competing Interests}
The authors declare no competing interests.

\section{Data Availability}
All code and data are available on OSF here: https://osf.io/395ea/

\bibliographystyle{apacite}
\bibliography{refs.bib}

\section{Supplemental Information}

In the main text, we reported using CRQA metrics with between-subjects similarity as a selection scheme for the model’s outputs. This selection scheme resulted in an error of roughly $M = 0.19$ out of a possible ceiling of $M\approx0.02$. To bring performance closer to ceiling, we consider other selection schemes by varying the selection metric and/or the style of similarity. In total, we considered 23 more selection schemes. None substantially reduced error.

We first tested using CRQA metrics and within-subjects similarity. Recall that in within-subjects similarity, the kth predicted timeseries is considered similar to the observed timeseries if both agent’s predicted timeseries are similar to their respective observed timeseries. This is quantified as the average between agent 1’s predicted-observed similarity and agent 2’s predicted-observed similarity. We defined similarity in terms of one of the four CRQA metrics (RR, DET, ENT, MaxL). The higher the value of one of these metrics, the higher the similarity. Using this selection scheme resulted in an error as low as $M = 0.18$ for the first quartile mean of MaxL.

Seeing that CRQA metrics with within-subjects similarity yields comparable errors as CRQA with between-subjects, we next considered different selection schemes altogether. We tried different between-subjects and within-subjects variants of various metrics, and also tested whether just the properties of the predicted timeseries itself, without comparison to any observed data, could be diagnostic.
\begin{itemize}
    \item Between-Subjects Similarity
    \begin{itemize}
        \item Correlation ($R$) between agents' predicted and agents' observed, then the difference between predicted and observed correlations
    \end{itemize}
    \item Within-Subjects Similarity
    \begin{itemize}
        \item Correlation ($R$) between an agent's predicted and observed
    \end{itemize}
    \begin{itemize}
        \item Mean squared difference (MSE) between an agent's predicted and observed timeseries
    \end{itemize}
    \begin{itemize}
        \item Coefficient of determination ($R^2$) between an agent's predicted and observed timeseries
    \end{itemize}
    \begin{itemize}
        \item Normalized sum of differences in timeseries derivatives (norm. sum derivative error), computed by taking the difference between adjacent timepoints in the predicted and the observed timeseries (timeseries derivatives), taking the sum of the absolute value of the pairwise differences between these derivatives, and dividing that sum by the range of the observed timeseries
    \end{itemize}
    \begin{itemize}
        \item The variance of differences in timeseries derivatives (variance derivative error), computed by taking the difference between adjacent timepoints in the predicted and the observed timeseries (timeseries derivatives), computing the variance in the pairwise differences between these derivatives, and dividing that variance by the range of the observed timeseries
    \end{itemize}
            \begin{itemize}
                \item Absolute difference in variances (variance difference) of an agent's predicted and observed timeseries
            \end{itemize}
            \begin{itemize}
                \item Normalized absolute difference in variances (norm. variance diff) of an agent's predicted and observed timeseries
                \item CRQA RR between an agent's predicted and observed timeseries
            \end{itemize}
            \begin{itemize}
                \item CRQA DET between an agent's predicted and observed timeseries
            \end{itemize}
            \begin{itemize}
                \item CRQA ENT between an agent's predicted and observed timeseries
            \end{itemize}
            \begin{itemize}
                \item CRQA MaxL between an agent's predicted and observed timeseries
            \end{itemize}
    \item Standalone features of the predicted timeseries:
    \begin{itemize}
        \item Variance
    \end{itemize}
    \begin{itemize}
        \item Mobility, one of the Hjorth parameters, computed as mean frequency
    \end{itemize}
    \begin{itemize}
        \item Complexity, another Hjorth parameter, computed as change in frequency
    \end{itemize}
    \begin{itemize}
        \item Smoothness, computed as the mean of the magnitudes of the predicted timeseries' first derivative
    \end{itemize}
    \begin{itemize}
        \item  Weighted smoothness, computed as the sum of absolute changes between adjacent timepoints in the predicted timeseries, scaled by the variance of the observed timeseries
    \end{itemize}
    \begin{itemize}
        \item Dominant frequency, the frequency with the highest magnitude in the Fourier transform (excluding the DC component)
    \end{itemize}
    \begin{itemize}
        \item Spectral entropy, the entropy of the normalized Fourier spectrum
    \end{itemize}
    \begin{itemize}
        \item Autocorrelation, the correlation between the series and itself shifted by one time step
    \end{itemize}
    \begin{itemize}
        \item Partial autocorrelation, the correlation between the series and itself one step apart after removing the linear effect of the intervening lags
    \end{itemize}
    \begin{itemize}
        \item Trend strength, the absolute slope of a linear fit to the timeseries
    \end{itemize}
    \begin{itemize}
        \item Number of peaks, the count of local maxima in the timeseries
    \end{itemize}
    \begin{itemize}
        \item Number of troughs, the count of local minima in the timeseries
    \end{itemize}
\end{itemize}

For the following within-subjects metrics, we chose the transition covariance whose predicted timeseries reflected the maximum metric value: correlation, $R^2$, norm. sum derivative error, RR, DET, ENT, and MaxL. For all other within-subjects metrics, we chose the minimum metric value. For standalone metrics, we selected the transition covariance based on either the maximum or minimum metric value, whichever yielded a smaller overall mean error based on the final matrix, as we did not have reason to prefer either extreme. 

\begin{figure}
\centering
\begin{subfigure}{0.8\textwidth}
    \centering
    \includegraphics[width=\textwidth]{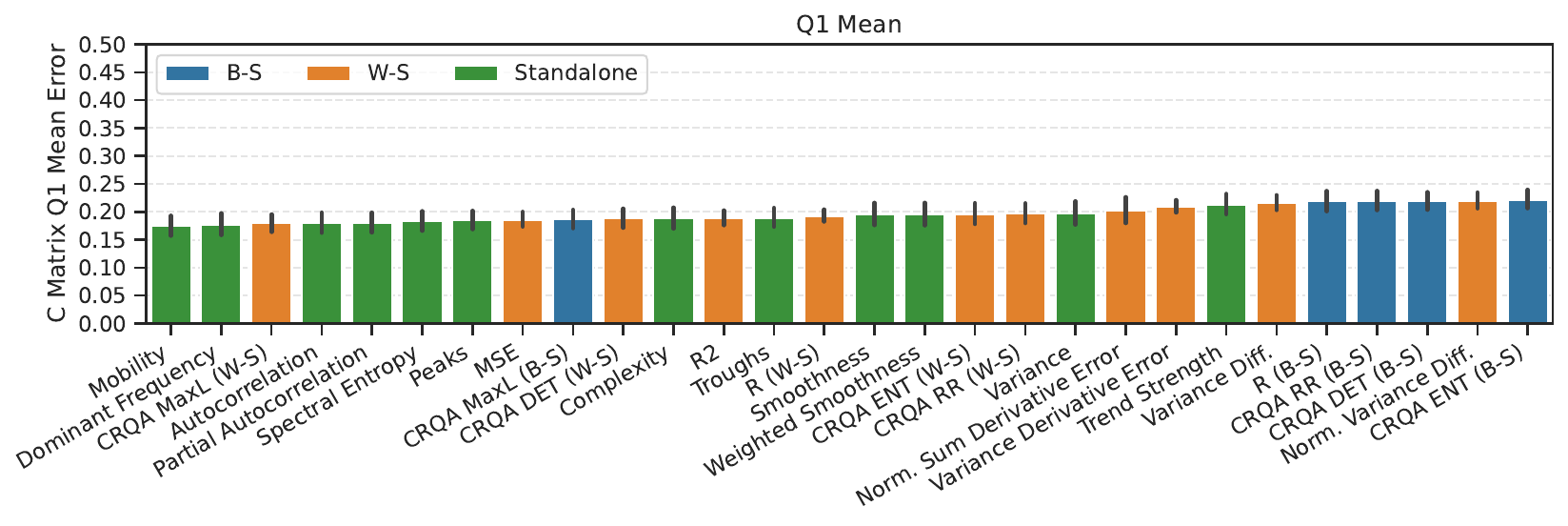}
    \caption{}
\end{subfigure}
\hfill
\begin{subfigure}{0.8\textwidth}
    \centering
    \includegraphics[width=\textwidth]{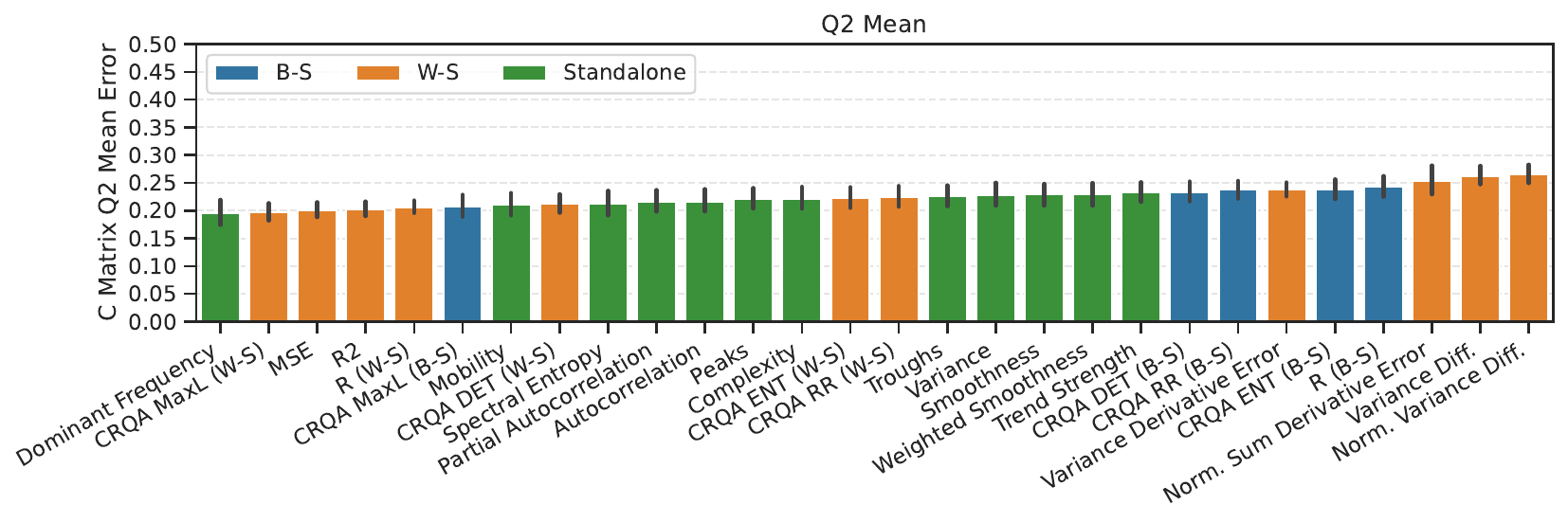}
    \caption{}
\end{subfigure}
\hfill
\begin{subfigure}{0.8\textwidth}
    \centering
    \includegraphics[width=\textwidth]{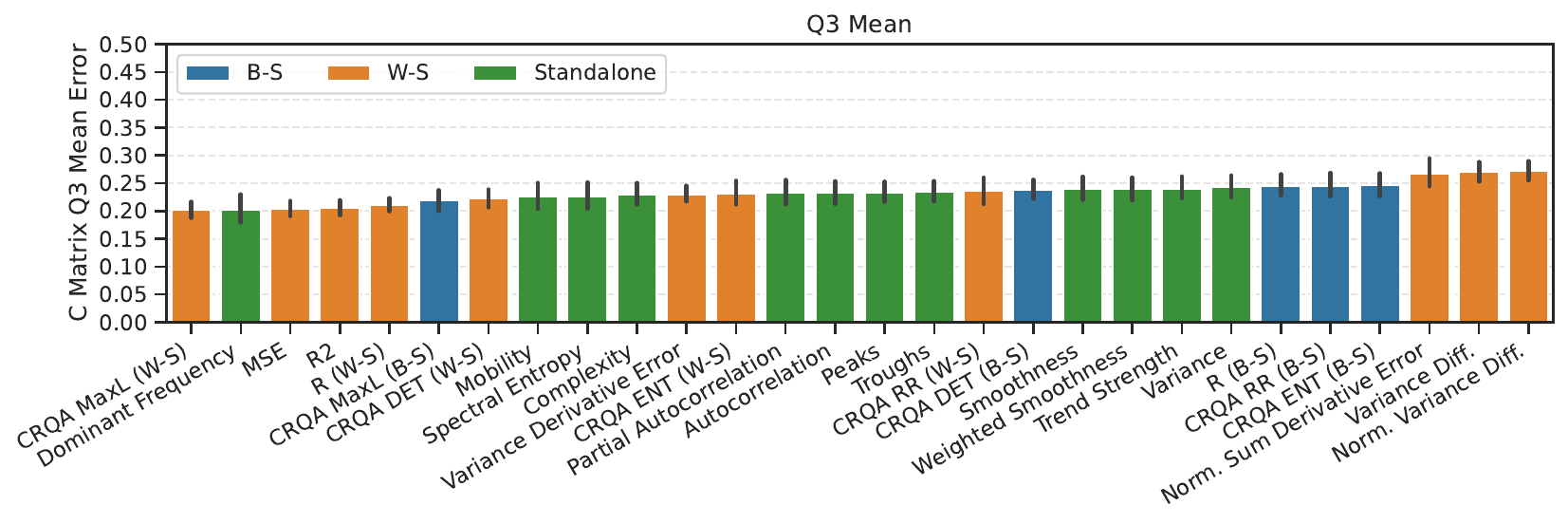}
    \caption{}
\end{subfigure}
\hfill
\begin{subfigure}{0.8\textwidth}
    \centering
    \includegraphics[width=\textwidth]{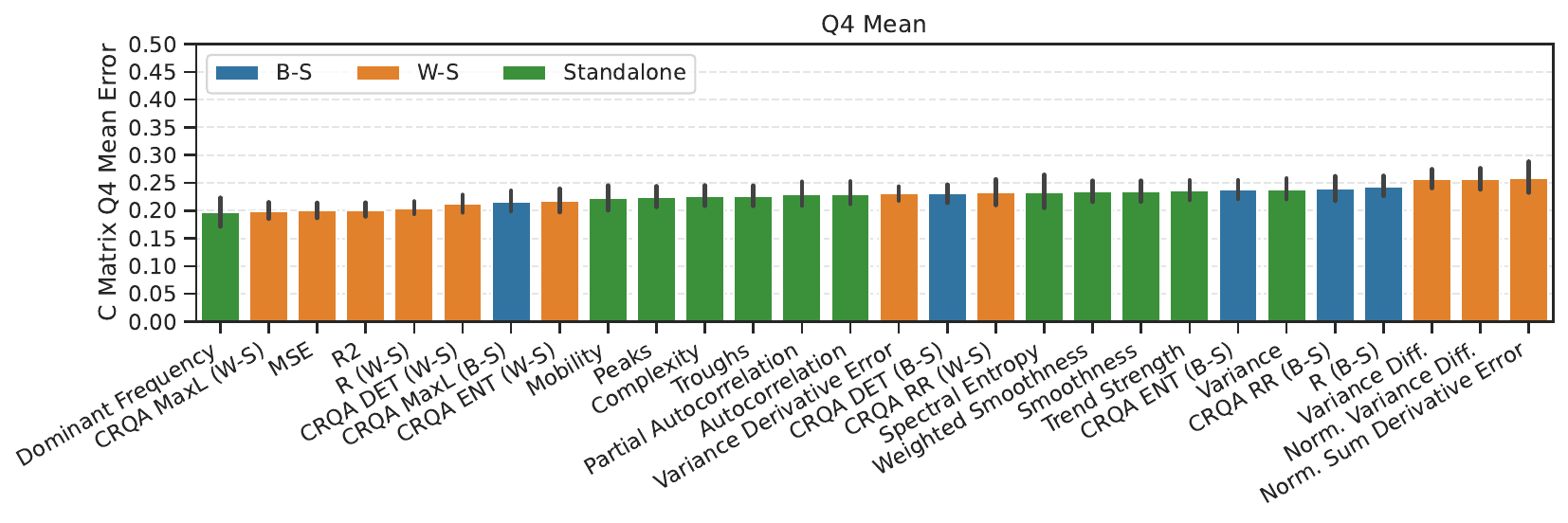}
    \caption{}
\end{subfigure}
\hfill
\begin{subfigure}{0.75\textwidth}
    \centering
    \includegraphics[width=\textwidth]{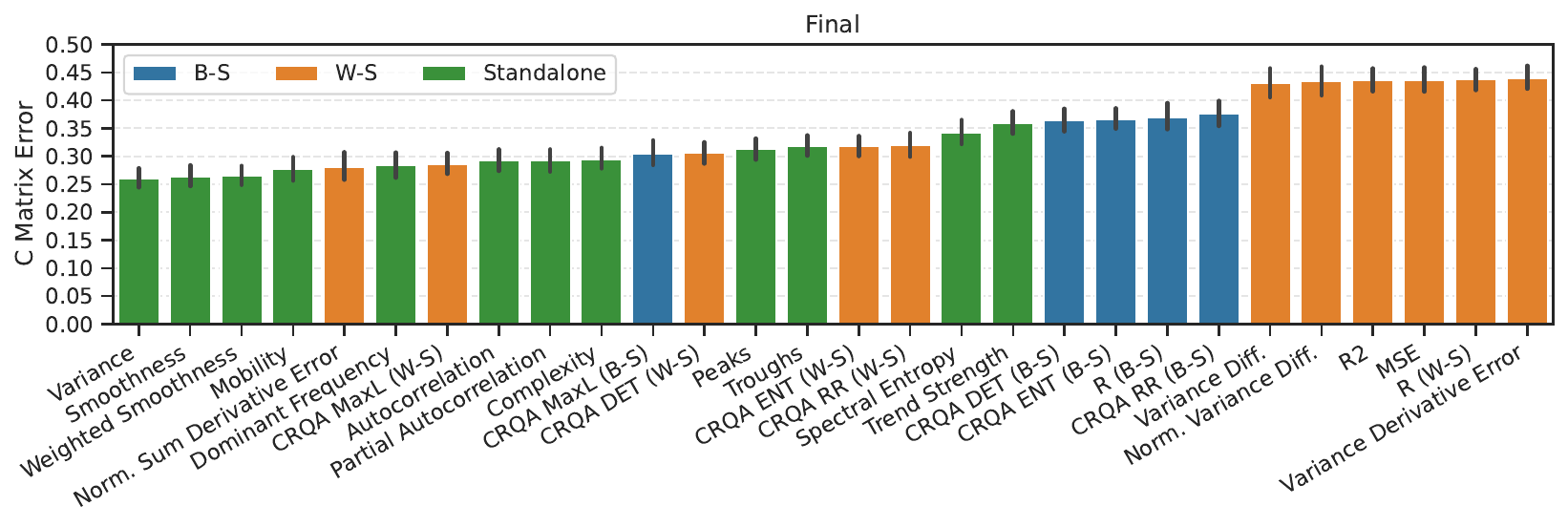}
    \caption{}
\end{subfigure}
\hfill
\caption{Context matrix errors for all metrics using (a) first quartile means, (b) second quartile means, (c) third quartile means, (d) fourth quartile means, and (e) the final matrix. Error bars are 95\% CIs.}
\label{fig:sup_figure}
\end{figure}

The errors using final matrix and quartile means for each of these selection schemes is shown in Figure~\ref{fig:sup_figure}. As shown, no selection scheme substantially reduces error beyond that of our selection schemes in the main text. The best selection scheme is mobility, a standalone property of the predicted timeseries, with $M = 0.17$ using the first quartile mean. 

\end{document}